\documentclass[journal=prl,aps,twocolumn]{revtex4-2} 

\usepackage{xr-hyper}   
\usepackage{orcidlink}

\usepackage{graphicx}  
\usepackage{dcolumn}   
\usepackage{bm}        
\usepackage{amssymb}   
\usepackage{hyperref}

\usepackage{natbib}

\usepackage{diagbox}

\usepackage{soul,color}

\usepackage{caption} 
\usepackage{subcaption} 

\usepackage{amsfonts}

\usepackage{amsmath}
\usepackage{blkarray}
\usepackage{braket}
\usepackage{multirow}
\usepackage{mathtools}

\usepackage{enumerate}

\usepackage{tikz}
\usepackage[utf8]{inputenc}

\usepackage{bm}

\newcolumntype{L}{>{$}l<{$}} 

\newcommand{\bk}{{\bf k}}
\newcommand{\bkp}{{\bf k'}}

\newcommand{\com}[1]{}

\DeclareMathOperator{\eV}{eV}

\newcommand{\rev}[1]{{\color{blue}{#1}}}  

\begin{document}

\title{
Fast Dipolar Charge-Polarization Fluctuations: Toward a Unified Pairing Mechanism in Bad Metals
}

\author{M. Ahsan Zeb ~\orcidlink{0000-0002-3448-2357}}
\affiliation{Department of Physics, Quaid-i-Azam University, Islamabad 45320, Pakistan}
\email{ahsan.zeb@qau.edu.pk}


\date{\today}
\begin{abstract}


The dynamics of a quantum system at low energies is strongly influenced by effective interactions generated by states at higher energies.
In 
effective models of correlated electron systems,
contributions
from the high energy bands
to the two-particle scattering vertex,
beyond dielectric screening,
are typically limited to
virtual hopping-induced purely kinetic corrections.
Considering a mixed kinetic-potential process 
involving the Coulomb interaction between particles
and hopping-induced fast dipolar charge-polarization fluctuations,
we obtain an analytical expression for the effective interaction between particles at low energies.
Applied to cuprates (doped Mott insulators) and iron-based systems (Hund's metals), 
 the interaction
yields the observed 
$d_{x^2-y^2}$ and $s^{+-}$ gap symmetries,
raising the prospect of a unified pairing mechanism across bad metals.


\end{abstract}

\maketitle

Understanding the pairing mechanism
in unconventional superconductors
is a central problem in condensed matter physics.
Unlike conventional superconductors---metals and alloys---where
 electron pairing is universally mediated by virtual phonons~\cite{BCS1957},
 a variety of class-specific pairing mechanisms 
 have been proposed for
 superconductivity
   in 
   strongly correlated electron systems~\cite{Scalapino2012,Stewart2017}.

For example, within the spin-fluctuation framework, 
superconductivity in cuprates
~\cite{LeeRMP2006,Keimer2015}
 with a Mott-insulating parent
is commonly attributed to local-moment 
superexchange~\cite{Scalapino2012}, 
whereas in iron-based superconductors~\cite{Paglione2010,Stewart2011,Fernandes2022}
 with
a metallic multiband parent, 
itinerant spin fluctuations, often associated with Fermi-surface nesting, 
are believed to mediate pairing~\cite{Mazin2008,Hirschfeld2011}. 
Alternative mechanisms based on orbital fluctuations 
and nematicity have also been proposed for 
the latter class~\cite{Kontani2010,Fernandes2014}.

Such correlated systems
are naturally described within a tight-binding framework. 
The low-energy effective lattice models of these systems
retain only the low-energy electronic orbitals and assume that
the role of the high-energy states is limited to
renormalizing the hopping parameters
and 
screening the Coulomb interaction~\cite{Aryasetiawan2004,Imada2010}.
To explain superconductivity,
the focus is primarily on
effective interactions \emph{arising within}
the low-energy model,
from kinetic processes (such as superexchange in cuprates)
or mediated by emergent collective spin, charge, or orbital excitations
(requiring proximity to the corresponding ordering instabilities)~\cite{Scalapino2012}.

The influence of the high energy states
beyond the roles described above
and pertinent to superconductivity
has received comparatively little attention.
In principle,
advanced numerical methods such as 
constraint functional renormalization group (cfRG)~\cite{Kinza2013}
can track, while downfolding,
all changes to 
 the two-particle scattering 
vertex $\Gamma=\Gamma({\mathbf{k}_1,\mathbf{k}_2,\mathbf{q}})$
(which describes the effective interaction
between two particles with momenta
 $\mathbf{k}_1$ and $\mathbf{k}_2$
 exchanging momentum $\mathbf{q}$).
Can the high energy states give rise to an analytically tractable pairing interaction?
More importantly, if yes,
can it account for the observed gap structures across seemingly distinct materials families?

Here, we address these questions and give an affirmative answer.
We note that the hopping between the low and the high energy orbitals
produces charge fluctuations, 
and the Coulomb interaction thereof 
with the other particles
should produce a contribution to 
$\Gamma$. 
Surprisingly, this simple correction has been overlooked thus far.

Considering a mean-field reference state, 
it is clear that such hopping processes induce dipolar charge fluctuations.
Due to the gap $\Delta_{\mathrm{CT}}$ between the two energy sectors,
these fluctuations occur on a much shorter timescale than
the low energy dynamics,
so they are only weakly screened~\cite{Ayral2013},
leading to a significant 
effect on
$\Gamma$.

It is more natural to consider the process in real space, where it mediates an effective interaction between the hopping particle and its neighbors, with a range that depends on the range
of the Coulomb interaction.
So under typical circumstances, 
it generates a long-range anisotropic
interaction that, as we shall see shortly,
can be \emph{attractive} in unconventional pairing channels.
Needless to say, 
 this mechanism does not require proximity to any 
  instability.

\begin{figure*}[htbp]
\begin{center}
\includegraphics[width=1\textwidth]{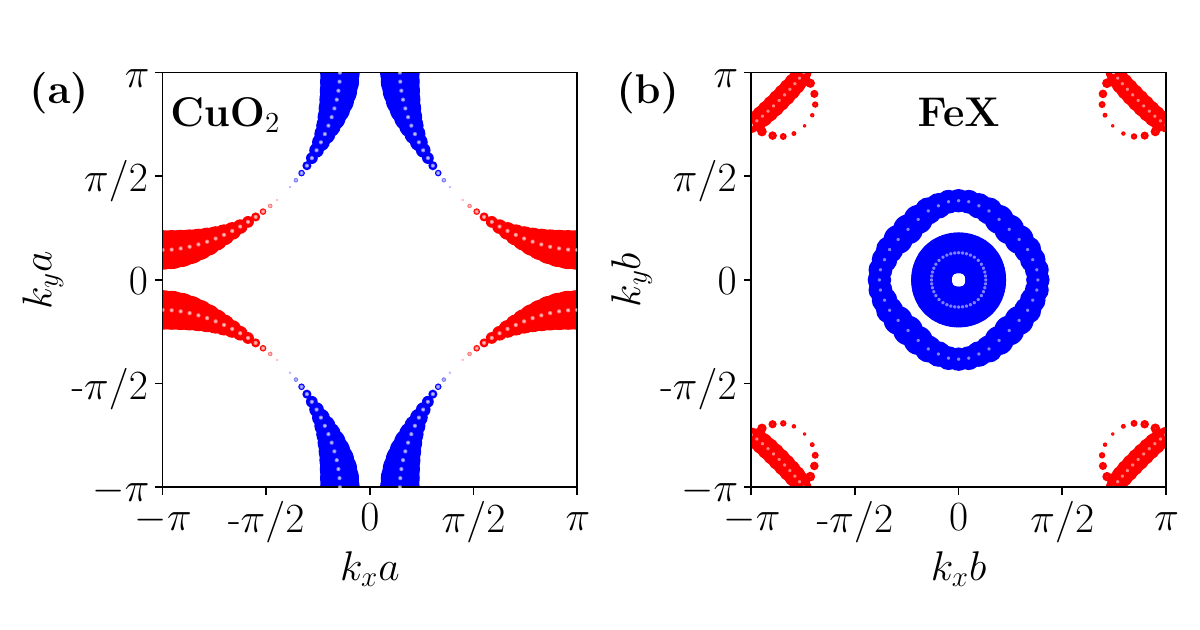}
\caption{
Superconducting gap functions with
$d_{x^2-y^2}$ and $s^{+-}$-like symmetries
given by the pairing interaction 
mediated by hopping-induced fast dipolar charge-polarization fluctuations
in CuO$_2$ (a) and Iron based superconductors (b).
Sign of the gap is shown by the color (red/blue) 
while its magnitude is proportional to the point/circle sizes.
The while dotted lines show the Fermi surface(s).
}
\label{fig:gaps}
\end{center}
\end{figure*}

We find that
in cuprates~\cite{LeeRMP2006,Keimer2015} and
iron-based superconductors~\cite{Paglione2010,Stewart2011,Fernandes2022},
this mechanism reproduces the experimentally observed
$d_{x^2-y^2}$ and $s^{+-}$ gap symmetries, respectively
(Fig.~\ref{fig:gaps}),
supporting the possibility of a common origin of superconductivity
in these two classes of correlated materials.

\begin{figure}[htbp]
\begin{center}
\includegraphics[width=1\columnwidth]{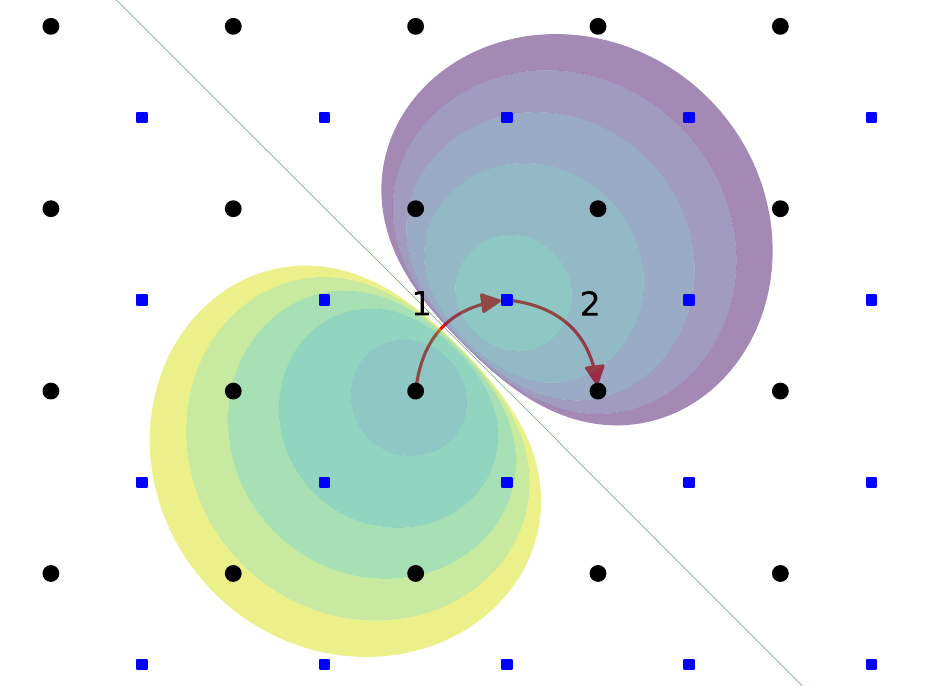}
\caption{Hopping induced dipolar charge-polarization fluctuations.
Particles in the low energy orbitals on the black lattice (circles) hop through auxiliary charge-gap orbitals on the blue lattice (squares), 
as indicated by the two arrows labeled $1$ and $2$.
An electric dipolar charge-polarization forms in the intermediate state,
which couples to the particles nearby via its Coulomb potential (shaded contours).
Fast fluctuations are weakly screened so they can mediate
an (anisotropic) interaction between particles 
that can overcome their Coulomb repulsion,
which is strongly screened at low energies.
}
\label{fig:generic}
\end{center}
\end{figure}

{\it Model}:
We demonstrate the idea using a 
 two-band model
of spinless electrons/holes, on a bipartite lattice for clarity.
The quadratic part is 
$H_{0} = \epsilon_{\mathrm{A}} \sum_{i} a^\dagger_{i} a_{i}
+ \epsilon_{\mathrm{B}}\sum_{j} b^\dagger_{j} b_{j}
+ t\sum_{\braket{ij}} \left(a^\dagger_{i} b_{j} + b^\dagger_{j} a_{i}\right) $,
where $a^\dagger_{i}$ ($b^\dagger_{i}$)
creates particles in a low (high) energy orbital
at $i$th site on sublattice A (B),
and $t$ is the nearest neighbor (NN) 
hopping amplitude
between the two 
sublattices.
Here, a charge transfer gap
$\Delta_{\mathrm{CT}}=|\epsilon_{\mathrm{B}}-\epsilon_{\mathrm{A}}|>0$
is assumed. 
The high energy auxiliary states can be 
completely filled states
deep below the low energy states ($\epsilon_{\mathrm{B}}<\epsilon_{\mathrm{A}}$) or
completely empty states well above them ($\epsilon_{\mathrm{B}}>\epsilon_{\mathrm{A}}$)
(or, alternatively, be strongly correlated for a gap to exist).
For concreteness, 
we take $\epsilon_{\mathrm{B}}>\epsilon_{\mathrm{A}}$ in the following.

Now, consider the Coulomb (density-density) interaction between the particles,
${H_{\mathrm{int}}=1/2\sum_{ij} v_{ij} n_{i}n_{j}}$,
where
${n_{i}\equiv n_{Ai}}$ ($n_{Bi}$)
is the number of particles at the $i$th site on sublattice A (B),
given by
${n_{Ai}=a_{i}^\dagger  a_{i}}$
 (${n_{Bi}=b_{i}^\dagger  b_{i}}$),
 and $v_{ij}$ is the strength of the \emph{bare} interaction.

By choosing a mean-field reference ground state 
of this many-particle system,
the full Hamiltonian $H=H_0+H_{\mathrm{int}}$
 can be written in terms of the corresponding mean-field Hamiltonian $H_{\mathrm{MF}}$ and fluctuations about it,
 as
 $H=H_{\mathrm{MF}} + \delta H_{\mathrm{slow}} + \delta H_{\mathrm{fast}}$,
 where we split the fluctuations based on their time-scale 
$\tau$
into \emph{slow} ($1/\tau\sim0$) and
 \emph{fast} ($1/\tau\sim\Delta_\mathrm{CT}$).
Since, in general, 
the dynamic dielectric screening 
is much stronger at low frequencies
(as ``the system'' has enough time to respond),
the Coulomb interaction in
the first two terms in $H$
should be much weaker 
than that in the last term, $\delta H_{\mathrm{fast}}$
(for which the system is too sluggish to act timely).

The low energy effective Hamiltonian is~\cite{ColemanBook}
$H_{\mathrm{eff}} = \mathcal{P} H \mathcal{P} + V_{\mathrm{eff}}$,
where $\mathcal{P}$ is projector onto the low energy space
and $V_{\mathrm{eff}}$ is 
a contribution to the two-particle scattering vertex
mediated by the states in the high energy space 
via the fast fluctuations $\delta H_{\mathrm{fast}}$.
To this end, 
beyond the dielectric screening,
the charge transfer arising from
 the inter-sublattice hopping term in $H_0$
 plays the key role, 
 so we take
$\delta H_{\mathrm{fast}} \approx \sum_{\braket{ij}, \ell} (W_{i\ell} \delta n_{Bi}+W_{j\ell} \delta n_{Aj}) n_{A\ell}$,
where
$W_{i\ell}$
 is the strength of the screened Coulomb repulsion (at $\omega=\Delta_\mathrm{CT}$),
 and
$\delta n_{Ai}$
 is the charge fluctuation at the $i$th site on sublattice A. 
Here, $\delta n_{Ai}=-\delta n_{Bi}=-1$ in units of elementary charge.
$V_{\text {eff}}$
can be easily computed using Schrieffer-Wolf transformations~\cite{Schrieffer1966}, and one obtains a novel 
nonlocal interaction
\begin{align}
\label{eq:Veff}
V_{\text {eff}} &= \frac{1}{2}
\sum_{\braket{i,jj'}\ell}
  V_{ij\ell}~
  a^\dagger_{j} a_{j'}
  a^\dagger_{\ell}a_{\ell},\\
  \label{eq:Vijl}
 V_{ij\ell} &= \frac{t^2}{\Delta_{\mathrm{CT}}^2}   \left(W_{i\ell} -W_{j\ell}\right),
\end{align}
where $\braket{i,jj'}$ shows $j,j'\in A$ to be neighbors of $i\in B$.

The pair potential for spin singlet pairing
 obtained from symmetrized Eq.~\ref{eq:Veff} is
$V(\mathbf{k,k'})=\sum_{\braket{i=0,jj'} \ell} V_{ijj'\ell} 
\cos\left(\mathbf{k}\cdot \boldsymbol\rho_{j\ell}\right)\cos\left(\mathbf{k'}\cdot  \boldsymbol\rho_{j'\ell} \right)$,
where
$V_{ijj'\ell}=(V_{ij\ell}+V_{ij'\ell})/2$ and
 $\boldsymbol\rho_{j\ell}=\mathbf{r}_{j}-\mathbf{r}_{\ell}$ 
 is the relative position of sites $\ell$ and $j$.
The key question is
whether, within Bardeen-Cooper-Schrieffer (BCS) theory~\cite{BCS1957},
 our model
leads to Cooper pairing with the correct superconducting gap symmetry.

\renewcommand{\arraystretch}{1.3}
\begin{table}[htp]
\caption{Mapping of cuprates and Iron based superconductors FeX (X=Se,Te,As, \ldots) onto our generic model.}
\label{tab:map}
\begin{center}
\begin{tabular}{|c|c|c|}
\hline
\diagbox{{\bf Quantity}}{{\bf  System}}
 & {\bf CuO$_2$} & {\bf FeX} \\ \hline
low energy states 
& ~O: $p_x,p_y$~ & Fe: $d_{xy}, d_{yz}, d_{zx}$\\ \hline
high energy states  
& ~Cu: $d_{x^2-y^2}$~ & X: $p_x,p_y,p_z$\\ \hline
charge gap $\Delta_{\mathrm{CT}}$ & $\epsilon_{p}-\epsilon_{d}$ & $\epsilon_{d}-\epsilon_{p}$ \\ \hline
\end{tabular}
\end{center}
\end{table}%

{\it Applications}: 
Hole-doped cuprates and iron pnictides/chalcogenides are two paradigmatic classes of unconventional superconductors,
where
the essential physics originates from the CuO$_2$ and FeX planes (X=Se,Te,As, \ldots), respectively.
Cuprates are effectively single-band systems~\cite{LeeRMP2006,Keimer2015}, 
whereas iron-based superconductors are multiband systems with multiple Fermi-surface pockets~\cite{Paglione2010,Stewart2011}.
Table~\ref{tab:map} identifies the low-energy itinerant and high-energy auxiliary states relevant to our model in these materials.
The pairing interaction in such muti-orbital cases is given in End Matters (EM). We discuss these cases in turn.

Hole-doped cuprates provide a natural test case, as doped holes predominantly reside on the O atoms~\cite{Zaanen1985,ChenPRL1991,Kowalski21}, which form the low-energy itinerant states that experience the effective attraction generated by our mechanism. We describe the electronic structure using 
Emery model~\cite{Emery1987},
which 
considers three orbitals, $d_{x^2-y^2}$ orbital of Cu
and $p_x$, $p_y$ orbitals of two O sitting along the two cartesian axes, respectively. 
The effective interaction is obtained from Cu-O nearest-neighbor hopping and intersite Coulomb interactions up to a lattice constant.
The symmetry of the superconducting gap function is obtained
by numerically solving the fermi-surface-projected linearized gap equation
for the most attractive eigenvector. 
We find that
$d_{x^2-y^2}$ 
is the dominant pairing channel~\cite{SM},
consistent with the observed cuprate gap symmetry~\cite{Tsuei2000}
[see Fig.~\ref{fig:gaps}(a)].
Including the
screened Coulomb repulsion
does not affect the results 
as long as 
$\xi\equiv 
t^2\varepsilon(0)/\Delta_{\mathrm{CT}}^2\varepsilon(\Delta_{\mathrm{CT}})$
(where $\varepsilon(\omega)$ is dielectric function),
which determines the relative strength of
the effective interaction, is large enough.
(At typical values of the parameters, $\xi_{th}\gtrsim 3.2$, which is well within physically relevant regime, see EM).
The applicability of our simple model improves with doping but is limited in the low-doping regime close to the parent Mott insulator, where the carriers are better described as Zhang--Rice singlets~\cite{ZhangRice1988} within the $t$-$J$ model~\cite{LeeRMP2006}.

Iron pnictides and chalcogenides are multiband systems with electron and hole pockets derived primarily from the Fe $t_{2g}$ orbitals ($d_{xy}$, $d_{yz}$, and $d_{zx}$)~\cite{Paglione2010,Stewart2011}. 
While a five-band model with one Fe per unit cell adequately describes the low-energy electronic structure~\cite{Kuroki2008,Graser2009},
the pairing interaction is evaluated in the crystallographic two-Fe unit cell to account for the two inequivalent X sites.
Accordingly, we derive~\cite{SM} a ten-orbital tight-binding model for FeSe from density-functional theory (\textsc{Quantum ESPRESSO}~\cite{Giannozzi2009}) 
using maximally localized Wannier functions~\cite{Marzari1997} (\textsc{Wannier90}~\cite{Mostofi2008,Pizzi2020})
The pairing interaction arises from virtual hopping between Fe $d$ orbitals via the $p$ orbitals of neighboring Se atoms, treated as completely filled auxiliary charge-transfer states (see EM for details). 
Including intersite Coulomb interactions up to $\sim 2a$ (fourth-neighbor Fe--Fe and second-neighbor Se--Fe), we solve the fermi-surface-projected linearized gap equation
and obtain an $s^{+-}$-like gap [Fig.~\ref{fig:gaps}(b)], with a sign change between the hole and electron pockets, weak anisotropy on most Fermi-surface sheets, oscillations on part of the hole pockets, and relative gap magnitudes consistent with experiment~\cite{Hirschfeld2011}.

{\it Discussion:}
Schrieffer-Wolff transformations, although valid only for $t/\Delta_{\mathrm{CT}}\ll1$, suggest that charge hopping generically generates effective interactions. In the opposite limit, $t/\Delta_{\mathrm{CT}}\gg1$, the low-energy states carry comparable spectral weight on both A and B sublattices,
 implying that dipolar charge-polarization fluctuations also become slow and are therefore strongly screened. 
 Since the probability $p=(t/\Delta_{\mathrm{CT}})^2$ (the spectral weight on the B sublattice) evolves continuously from the perturbative limit to $1/2$, the effective interaction $V_{\mathrm{eff}}$ may remain significant over part of the crossover.

Our model may also be relevant to the strange-metal phase, which emerges above the superconducting dome in the phase diagrams of many correlated materials~\cite{Bruin2013,Keimer2015}.
Its $T$-linear resistivity requires an efficient mechanism for momentum relaxation, e.g., through umklapp scattering.
A prominent class of theories attributes this behavior to coupling between low-energy quasiparticles and a critically damped emergent bosonic mode~\cite{Abanov2003}.
Our preliminary analysis suggests that umklapp scattering due to 
$V_{\mathrm{eff}}$ is strong for quasiparticle pairs over much of the Fermi surface, limited primarily by the available phase space. 
Whether this mechanism can account for the observed $T$-linear resistivity remains an interesting question for future work.

Combined with existing first-principles downfolding workflows and many-body frameworks, the effective interaction derived here can be evaluated using realistic hopping and screened intersite Coulomb matrix elements, both extending beyond nearest neighbors, enabling (more accurate) material-specific calculations.

{\it Conclusion:}
We have shown that hopping-induced fast dipolar charge-polarization fluctuations generate a previously unrecognized effective interaction. 
Its success in reproducing the pairing symmetries of cuprates and iron-based superconductors suggests that high-energy charge dynamics may provide a common microscopic ingredient of superconductivity in bad metals.

\bibliographystyle{apsrev4-2}
\bibliography{perovskites}

\section*{End Matters}

\appendix

\subsection{Pairing interaction in multi-orbital cases}

Considering multiple atoms/orbitals per unit cell,
the labels expand to a pair of cell index and a composite atom/orbital index, as 
$i,\alpha$.
For this multi-atom multi-orbital case, 
Eqs.~\ref{eq:Veff}-\ref{eq:Vijl} are updated to
\begin{gather}
\label{eqEM:Veff}
V_{\text {eff}} = \frac{1}{2}
\sum_{ijj'\ell}
\sum_{\braket{\alpha,\beta\beta'}\gamma}
  V_{ij\ell,\alpha\gamma}^{\beta\beta'}~
  a^\dagger_{j \beta} a_{j' \beta'}
  a^\dagger_{\ell \gamma}a_{\ell \gamma},\\ \nonumber
V_{ij\ell,\alpha\gamma}^{\beta\beta'}
=  \frac{T_{j\beta j'\beta'}^{i\alpha} }{\Delta_{\mathrm{CT}}}\left(W_{i\alpha,\ell\gamma} -W_{j\beta,\ell\gamma}\right),\\ \nonumber
T_{j\beta j'\beta'}^{i\alpha} 
=
\frac{ t_{i\alpha,j\beta}^* t_{i\alpha,j'\beta'} }
{\Delta_{\mathrm{CT}} },
\end{gather}
where $\alpha$ label the auxiliary atoms/orbitals,
while
$T_{j\beta j'\beta'}^{i\alpha} $ are effective hopping amplitudes between $j\beta$ and $j'\beta'$ main orbitals 
(creation operators $a^\dagger_{j \beta}$)
mediated by the orbital $i\alpha$.
The first sum is severely restricted
by the NN condition $\braket{\alpha,\beta\beta'}$
in the second sum.
$t_{i\alpha,j\beta}$ are the hopping amplitudes 
for neighboring atoms.
For simplicity, here
we assumed a single charge transfer gap $\Delta_{\mathrm{CT}}$ for all orbitals. 

Transforming to momentum space,
symmetrized Eq.~\ref{eqEM:Veff} 
in the pairing approximation
becomes
\begin{multline}
\label{eqEM:Vkkp}
V_{\text {eff}} =
\frac{1}{2N}\sum_{\bk\bkp\sigma} 
\sum_{\beta\beta'\gamma}
V_{\beta\beta'\gamma\gamma}(\mathbf{k,k'})\\ \times
  a^\dagger_{\bk\sigma \gamma} 
  a^\dagger_{-\bk\bar\sigma \beta} a_{-\bk'\bar\sigma \beta'}
  a_{\bkp\sigma \gamma},
\end{multline}
where
\begin{multline*}
V_{\beta\beta'\gamma\gamma}(\mathbf{k,k'})=
\sum_{i=0,jj'\ell}
\sum_{\braket{\alpha}_{\beta\beta'}\gamma}
V_{ijj'\ell,\alpha\gamma}^{\beta\beta'} \\ \times
\cos\left(\mathbf{k}\cdot \boldsymbol\rho_{j\beta\ell\gamma}\right)\cos\left(\mathbf{k'}\cdot  \boldsymbol\rho_{j'\beta'\ell\gamma} \right)
\end{multline*}
is the pair potential for spin singlet pairing,
with 
$V_{ijj'\ell,\alpha\gamma}^{\beta\beta'}=(V_{ij\ell,\alpha\gamma}^{\beta\beta'} + V_{ij'\ell,\alpha\gamma}^{\beta\beta'})/2$,
 $\boldsymbol\rho_{j\beta\ell\gamma}=\mathbf{r}_{j\beta}-\mathbf{r}_{\ell\gamma}$ 
 being the relative position of sites $\ell\gamma$ and $j\beta$,
and the notation $\braket{\alpha}_{\beta\beta'}$ indicating that $\alpha$ has to be the common NN of both $\beta,\beta'$ atoms/orbitals.
We have also included 
 the spin indices 
${\sigma,\bar\sigma}$ 
(${\bar\sigma=-\sigma}$) for clarity and completeness.

The effective pairing interaction between Bloch states at the fermi surface(s)
is obtained by a unitary transformation of Eq.~\ref{eqEM:Vkkp},
which employs 
selected eigenvectors of 
the tight binding model.

\subsection{Cuprates}

{\it Pair potential:}
Eq.~\ref{eqEM:Vkkp} gives the pair potential
with the following substitutions.
For two O sublattices $X,Y$,
ignoring the spin,
$a_{j \beta} \to p_{j,x/y}$ 
with Fourier transforms
$a_{\bk \beta}\to p_{\bk,x/y}$.
$t_{i\alpha,j\beta}=\pm t_{pd}$ are the Cu-O hopping amplitudes.

{\it Parameters:}
To obtain the non-interacting hole bands and the fermi surface,
the Emery model is solved at
$\{\epsilon_d,\epsilon_p,t_{pd},t_{pp}\}=\{0,1.6,1.3,0.6\}\eV$.
Assuming that the parameters for the screened Coulomb interaction 
obey 
$W_{\alpha\beta}(0)\varepsilon(0)
=W_{\alpha\beta}(\Delta_{\mathrm{CT}}) \varepsilon(\Delta_{\mathrm{CT}})$,
a single parameter
$\xi\equiv 
t_{pd}^2\varepsilon(0)/\Delta_{\mathrm{CT}}^2\varepsilon(\Delta_{\mathrm{CT}})$,
can be used with statically screened intersite parameters $W_{\alpha\beta}(0)$:
$V_{pd},V_{pp},V_{pp}'$,
where
$V_{pd}$ is the intersite repulsion between NN Cu-O,
and $V_{pp},V_{pp}'$ are intersite repulsions for first and second NN O-O.

At ${V_{pd}:V_{pp}:V_{pp}'=1:0.7:0.5}$,
we find a competition between 
$B_{1g}$ and $B_{2g}$,
where $B_{1g}$ is dominant
for $\xi>\xi_{\mathrm{th}}\simeq2.5$.
Including the onsite Hubbard repulsion,
${U_d:U_p=7:1.5}$
 for Cu and O orbitals, shifts
  $\xi_{\mathrm{th}}$ slightly upwards, to $3.2$.

\subsection{Iron based systems}

Eq.~\ref{eqEM:Vkkp} gives the pair potential
with $a_{\bk \beta} \to d_{\bk \beta}$,
where $d_{\bk \beta}$
are Fourier transforms of d-orbitals of the two Fe atoms
while the auxiliary orbitals
$\alpha$ are $p$-orbitals of Se atoms.
$t_{i\alpha,j\beta}$ are the hopping amplitudes 
for neighboring Fe and Se atoms,
which are calculated using the standard Slater-Koster method,
from two Slater-Koster parameters, $pd\sigma, pd\pi$.

In numerical calculations of pairing interaction,
we only consider the $t_{2g}$ orbitals ($d_{xy}$, $d_{xz}$ and $d_{yz}$) on Fe sites, which contribute the most to the states at the fermi surfaces.
This also favors our assumption of a single charge transfer gap $\Delta_{\mathrm{CT}}$
as the $e_{g}$ orbitals have a relatively different onsite energy
due to crystal field effects.


{\it Parameters:}
To obtain the symmetry of 
the gap function, 
only the ratio
$pd\pi/pd\sigma$
of the 
Slater-Koster parameters needs to be specified.
For the same purpose, 
 the charge gap $\Delta_\mathrm{CT}$ and 
 one of the Coulomb parameters 
 are also eliminated in setting the overall normalization.
We used $pd\pi/pd\sigma=0.3$,
and Coulomb parameters
$V_{dd}:V_{dd}':V_{dd}'':V_{dd}'''=1:0.6:0.3:0.2$ for first to fourth NNs Fe-Fe
and
$V_{pd}:V_{pd}'=1.1:0.4$ for first and second NNs Se-Fe,
roughly in accordance with the intersite separations.

\end{document}


\title{
Supplemental Material: Fast Dipolar Charge-Polarization Fluctuations: Toward a Unified Pairing Mechanism in Bad Metals
}

\author{M. A. Zeb ~\orcidlink{0000-0002-3448-2357}}
\affiliation{Department of Physics, Quaid-i-Azam University, Islamabad 45320, Pakistan}
\email{ahsan.zeb@qau.edu.pk}


\date{\today}

\maketitle

\section{Derivation of effective interaction}

Consider the auxiliary states lying far below the low energy states,
and thus completely filled,
$\braket{b^\dagger_{i} b_{i}}=1$ $\forall i$.
Particle from these orbitals can hop to the neighboring itinerant lattice sites/orbitals (with amplitudes $t_{ij}$ for neighboring sites $i,j$).
The generator of the 
Schrieffer-Wolff transformation~\cite{Schrieffer1966}
in this case is simply
$S = S_{-} - S_{-}^\dagger$,
$S_{-} 
= \sum_{\braket{ij}} 
\frac{t_{ij} }{\epsilon_{\mathrm{A}} - \epsilon_{\mathrm{B}}}
  a^\dagger_{j} b_{i}$.
Here, the denominator contains the charge transfer gap,
${\epsilon_{\mathrm{A}} - \epsilon_{\mathrm{B}}= \Delta_{\mathrm{CT}} >0}$.
The low energy effective model is
$H_{\text {eff}}=\mathcal P e^{S}He^{-S} \mathcal P$,
where $H$ is the full Hamiltonian and $\mathcal P$ projects onto the low energy space,
which gives  
the effective interaction 
$V_{\text {eff}} = \frac{1}{2}\mathcal P [S,[S, \hcf] \mathcal P=S_{-}^\dagger \hcf S_{-}$.
This expression simplifies to Eq.~\ref{eq:Veff} in the Letter.
If we take the auxiliary states to lie well above the main states and be completely empty, $\braket{b^\dagger_{i} b_{i}}=0$ $\forall i$, then
$V_{\text {eff}} = S_{-} \hcf S_{-}^\dagger$,
which also leads to the same final expression. 

\com{
\subsection{Strongly correlated half-filled auxiliary states}
If the auxiliary states are half-filled correlated states (Mott insulators),
both emptying them or doubly occupying them will cost energy.
We include spin (labeled by $\sigma$) in this case, and
following Schrieffer and Wolff~\cite{Schrieffer1966},
we define
$S = \sum_{\xi=\pm} S_{\xi} - S_{\xi}^\dagger$,
$S_{\xi} 
= \sum_{\braket{ij}\sigma} 
\frac{t_{ij} }{\epsilon_{\text{A}} - \epsilon_\xi}
  n_{Bi-\sigma}^{\xi} a^\dagger_{j\sigma} b_{i\sigma}$,
where 
$\epsilon_\pm$ are energies of doubly occupied ($\xi=+$) and empty ($\xi=-$)
auxiliary correlated site,
and $n_{Bi\sigma}^{\pm} = n_{Bi\sigma}, 1-n_{Bi\sigma}$,
Addition or removal of a particle at an auxiliary (correlated) site
excites the system to higher energies,
costing Hubbard repulsion $U$,
or charge transfer energy $\Delta_{\text{CT}}$.
With these two types of excited subspaces 
$\delta n_i=\pm 1$ for a specific correlated site $i$,
and one of its nearest neighboring main (itinerant) site $j$, $\delta n_j=-\delta n_i$.
In this case, 
$V_{\text {eff}}=
S_{-}^\dagger \hcf S_{-} + S_{+} \hcf S_{+}^\dagger$,
and using the Pauli-Fierz identity,
we obtain
\begin{align}
V_{\text {eff}} &= \frac{1}{2}
\sum_{\braket{i,jj'}\ell} 
  V_{ijj'\ell}~
  c^\dagger_{j\sigma} c_{j'\sigma}
  c^\dagger_{\ell\sigma'}c_{\ell\sigma'},\\
 V_{ijj'\ell} &= t_{ij} t_{ij'}^*   \left( \frac{W_{i\ell}}{\Delta_{CT}^2} + \frac{W_{i\ell}}{U^2}  - \frac{W_{j\ell}}{\Delta_{CT}^2} - \frac{W_{j'\ell}}{U^2}\right),
\end{align}
where 
 a sum over repeated spin indices is implied,
 and we have dropped a term that describes 
the scattering off the Kondo-like singlet.
 It reduces to the same form as Eq.~\ref{eq:Veff}
 in the infinite $U$ limit. 
}

\section{Iron based systems}


We take FeSe as a reference system for FeX systems.
It crystallizes in a
tetragonal PbO-type crystal structure, space group P4/nmm~\cite{Hsu2008}.
The lattice parameters we consider 
are $a=b=3.93$~\AA, $c=5.85$~\AA.

\subsection{Density functional theory calculations}
LDA+U 
density functional theory calculations
are performed using 
Quantum Espresso (QE) package~\cite{Giannozzi2009},
with
Perdew Zunger~\cite{Perdew1981} flavour of 
the local density approximation,
and pseudopotentials for Fe and Se
taken from QE library.
Hubbard U and Hund's J parameters for Fe d-orbitals 
are taken as $4.3~\eV$ and $1.0~\eV$.
The plane wave cutoffs used for wavefunctions and density were $60$~Ry and $480$~Ry, while a ${4\times4\times4}$ mesh is used for the Brillouin zone integrations.
For self-consistency, 
the convergence threshold of $10^{-5}$~Ry is set on the total energy.

\subsection{Maximally localized Wannier functions}
A tight binding model describing only 
the low energy bands 
is obtained using the Wannier90 package~\cite{Mostofi2008,Pizzi2020}.
This downfolding is performed
for $10$ lowest energy bands around the fermi level
to calculate corresponding
maximally localized Wannier functions~\cite{Marzari1997} with $d$-orbital symmetry 
located at the Fe sites, and a $10$-orbital tight binding model
in their bases.

\com{The first and second NNs of Fe are at a distance
of $a/\sqrt{2}\simeq 0.7$ and $a=1$,
whereas, 
at $z=\pm 0.2576 c$ (z-coordinate of Se atoms relative to the Fe atom layer), 
the first and second NNs of Se are at a distance of
$\sqrt{(a/2)^2+z^2}\simeq 0.63$ and $\sqrt{(a/2)^2+a^2+z^2}\simeq 1.2$.
}

\subsection{Gauge choice for the tight binding model}

Care must be taken to consider a consistent gauge choice for the tight binding Hamiltonian eigenstates and the pairing interaction formula.
We choose the usual gauge employed in analytical tight binding calculations, where phase factors for the orbital/atomic positions (Wannier centers)
are considered. 
We used the python module \texttt{tbmodels}~\cite{tbmodels}
for the tight binding calculations, where 
this gauge can be chosen
by setting \texttt{convention=1} as an input to 
\texttt{hamilton()} function that builds the Hamiltonian 
at a given k-point in the Brillouin zone.
This is not included in the documentation of the package 
but can be learnt from the relevant source file \texttt{\_tb\_model.py}.

\bibliography{perovskites}